\documentclass[12pt]{article}
\usepackage{amsmath,amsfonts}
\usepackage{bm}
\def\be{\begin{equation}}
\def\ee{\end{equation}}
\def\bea{\begin{eqnarray}}
\def\eea{\end{eqnarray}}

\begin{document}

\title{Non-abelian $D=11$ Supermembrane.}
\author{\text{ M.P. Garc\'{\i}a del Moral$^a$\dag\quad and A. Restuccia \ddag}\\
\setlength{\baselineskip}{6pt}
 {\small \dag Departamento de F\'{\i}sica, Universidad de Oviedo;}\\
{\small c/ Calvo Sotelo 18, 33007, Oviedo, Spain.}\\
{\small \ddag Departamento de F\'\i sica, Universidad Sim\'on Bol\'\i var}\\
{\small Apartado 89000, Caracas 1080-A, Venezuela}\\
\footnote{\em e-mail: garciamormaria@uniovi.es, arestu@usb.ve}  }
\maketitle
\ \vskip -4in {\em \
\parbox[t]{1.0\linewidth}{\small\rightline{FFUOV-09/10} }
}

\vskip 3.1in

\begin{abstract}
We obtain a $U(M)$ action for supermembranes
 with central charges in the Light Cone Gauge (LCG).
The theory
realizes all of the symmetries and constraints of the supermembrane
together with the invariance under a U(M)
gauge group with $M$ arbitrary. The worldvolume action has
 (LCG) $\mathcal{N}=8$ supersymmetry and it corresponds to $M$ parallel
supermembranes minimally immersed on the target $M_9$x$T^2$ (MIM2).
In order to ensure the invariance under the symmetries and to close the
corresponding
algebra,  a star-product  determined by the central charge condition is introduced.
It is constructed with a nonconstant symplectic two-form where curvature terms
 are also present. The theory is in the strongly coupled gauge-gravity regime.
At low energies, the theory enters in a decoupling limit and it is described by an
ordinary $\mathcal{N}=8$ SYM in the IR phase for any number of M2-branes. \end{abstract}

\setlength{\parindent}{0cm}

\setlength{\baselineskip}{15pt}

\setlength{\parindent}{.6cm}

\setlength{\baselineskip}{15pt}


\section{Introduction}

An ultimate goal of String theory is to find its nonperturbative
quantization. M-theory has been elusive to this point although
significative advances have been realized. To get contact with four
dimensions, at the end of the day a phenomenological model that will
take into account this nonperturbative effects will be required.
From that point of view, obtaining  a non-abelian gauge formulation
directly from M-theory  -not just in its effective action- but in
the full-fledged formulation is an important goal. In this letter we
summarize the results obtained in \cite{GarciadelMoral:2009uf} where
we have been able to find a non abelian extension of the
supermembrane minimally immersed in a $M_9\times T^2$.

The supermembrane with a topological restriction associated to
an irreducible winding has been shown to have very interesting
properties: discreteness of the supersymmetric spectrum \cite{Boulton:2002br, Boulton:2004yt,Boulton:2006mm}, spontaneous
breaking of supersymmetry, stabilization of most of the moduli \cite{GarciadelMoral:2007xj}, a spectrum containing dyonic
strings plus pure supermembrane excitations \cite{GarciadelMoral:2008qe},
formulation on a G2 manifold \cite{Belhaj:2008qz}.
This restriction can be seen at algebraic level as  a central charge
condition on the 11D  supersymmetric algebra and geometrically as a
condition of being minimally immersed into the target space \cite{Bellorin:2005kj}, so from
now on, we will denote it as MIM2. The semiclassical supermembrane subject to an irreducible wrapping was first analyzed in \cite{Duff:1987cs}.\newline

Recently, the low energy conformal description of multiple M2-branes have received a lot of attention from the scientific community.
The original motivation has been to realize Maldacena's
Conjecture for M-theory \cite{Maldacena:1997re} according to which
M-theory/$AdS_4\times S^7$ should be dual to a $CFT_3$ generated by
the action of multiple M2-branes in the decoupling limit, that is ,
for a large number $M$ of M2's. \cite{Bagger:2006sk},\cite{Bagger:2007jr} and independently \cite{Gustavsson:2007vu} were the
first to obtain a realization of this algebra by imposing fields
to be evaluated on a three algebra with positive inner metric. In terms of a unique finite dimensional gauge group $SO(4)$
 with a twisted
Chern-Simons terms. In order to generalize it for
general $SU(M)$ gauge groups, a Chern-Simons-matter
theory  with $\mathcal{N}=6$  was found by \cite{Aharony:2008ug} (ABJM) in which they are able to
generalize the theory to an arbitrary $SU(N)$ and recover also BLG
theory for the case of $N=2$. The ABJM, or at least a sector of it,
can be also recover from the 3-algebra formulation by relaxing the
condition of total antisymmetry of the structure constants
\cite{Bagger:2008se}. A more complete list of relevant references for interested reader can be found in \cite{GarciadelMoral:2009uf}.
\newline

In this letter we
summarize the results obtained in \cite{GarciadelMoral:2009uf}. The approach is very different to the ABJM one.
 In  \cite{GarciadelMoral:2009uf} it is consistently extended  the action of a single MIM2 to a theory of
interacting parallel M2-branes minimally immersed (MIM2's) preserving all of the symmetries of the theory:
supersymmetry and invariance under area preserving
 diffeomorphisms. The theory is not conformal invariant. In the extension, the
  gauge and gravity sectors are strongly correlated. It corresponds
to have a M-theory dual of the Non-Abelian Born-Infeld action
describing a bundle of multiple D2-D0 branes, so  we work in the high energy
approximation. When the energy scale is low, the theory decouples and
it is effectively described by a $\mathcal{N}=8$ SYM in the IR phase.
As the energy scale raises the YM coupling constant becomes weaker and at some point,
oscillations modes of the pure supermembrane appear and the theory enters in the strong correlated gauge-gravity sector.\newline

\section{A Supermembrane with discrete spectrum:\\ the MIM2}
In this section we will make a self-contained summary of the
construction of the minimally immersed M2-brane (MIM2). The
hamiltonian of the $D=11$ Supermembrane \cite{Bergshoeff:1987cm} may
be defined in terms of maps $X^{M}$, $M=0,\dots, 10$, from a base
manifold $R\times \Sigma$, where $\Sigma$ is a Riemann surface of
genus $g$ onto a target manifold which we will assume to be 11D
Minkowski. The canonical reduced hamiltonian to the light-cone gauge
has the expression \cite{deWit:1988ig}

\begin{equation}\label{equ1}
  \mathcal{H}= \int_\Sigma  d\sigma^{2}\sqrt{W} \left(\frac{1}{2}
\left(\frac{P_M}{\sqrt{W}}\right)^2 +\frac{1}{4} \{X^M,X^N\}^2-
\overline{\Psi}\Gamma_{-}\Gamma_{M}\{X^M,\Psi\}\right)
\end{equation}
subject to the constraints \begin{equation}  \label{e2}
\phi_{1}:=d(\frac{P_{M}}{\sqrt{W}}dX^{M} -\overline{\Psi}\Gamma_{-}d\Psi)=0 \end{equation} and
\begin{equation} \label{e3}
\phi_{2}:=
   \oint_{C_{s}}(\frac{P_M}{\sqrt{W}}dX^M -\overline{\Psi}\Gamma_{-}d\Psi)= 0,
\end{equation}
where the range of $M$ is now $M=1,\dots,9$ corresponding to the
transverse coordinates in the light-cone gauge, $C_{s}$,
$s=1,\dots,2g$ is a basis of  1-dimensional homology on $\Sigma$,
 \be \label{e4}\{X^{M}, X^{N}\}= \frac{\epsilon
^{ab}}{\sqrt{W(\sigma)}}\partial_{a}X^{M}\partial_{b}X^{N}. \ee
$a,b=1,2$ and $\sigma^{a}$ are local coordinates over $\Sigma$.
$W(\sigma)$
 is a scalar density introduced in the light-cone gauge fixing procedure.
 $\phi_{1}$ and $\phi_{2}$ are generators of area preserving diffeomorphisms, see \cite{deWit:1989vb}. That is
\be \sigma\to\sigma^{'}\quad\to\quad W^{'}(\sigma)=
W(\sigma).\nonumber \ee When the target manifold is simply connected
$dX^{M}$ are exact one-forms.

The spectral properties of (\ref{equ1}) were obtained in the context of a
 $SU(N)$ regularized model \cite{deWit:1988ig} and it was shown to
have continuous spectrum from $[0,\infty)$.

This property of the theory relies on two basic facts: supersymmetry
and the presence of classical singular configurations, string-like
spikes, which may appear or disappear without changing the energy of
the model but may change the topology of the world-volume. Under
compactification of the target manifold generically the same basic
properties are also present and consequently the spectrum should be
also continuous \cite{deWit:1997zq}. In what follows we will impose a
topological restriction on the configuration space. It characterizes
a $D=11$ supermembrane with non-trivial central charges generated by
the wrapping on the compact sector of the target space
\cite{GarciadelMoral:2001zb},\cite{Boulton:2001gz},\cite{Boulton:2002br},\cite{Boulton:2006mm}.
We will consider in this paper the case $g=1$ Riemann
surface as a base manifold $\Sigma$ on a $M_9$x$T^2$ target space.
The configuration maps satisfy:

\begin{equation}\label{e5}
 \oint_{c_{s}}dX^{r}=2\pi L_{s}^{r}R^{r}\quad r,s=1,2.
\end{equation}\begin{equation}\label{e7}
 \oint_{c_{s}}dX^{m}=0 \quad m=3,\dots,9
\end{equation}
\\

 where $L^{r}_{s}$ are integers and $R^{r}, r=1,2$ are the radius of $T^{2}$.
  This conditions ensure that we are mapping $\Sigma$ onto a
$T^2$ sector of the target manifold.

We now impose the central charge condition \be\label{e8}
I^{rs}\equiv \int_{\Sigma}dX^{r}\wedge dX^{s}=(2\pi
R_{1}R_{2})n\epsilon^{rs} \ee where $\omega^{rs}$ is a symplectic
matrix on the $T^{2}$ sector of the target and $n=det L_i^r$ represents the irreducible winding.

 The topological condition
(\ref{e8}) does not change the field equations of the hamiltonian
(\ref{equ1}).
  In fact, any variation of $I^{rs}$ under a change $\delta X^{r}$, single valued over
 $\Sigma$, is identically zero. In addition to the field equations obtained from (\ref{equ1}),
the classical configurations must satisfy the condition (\ref{e8}).
It is only a topological restriction on the original set of
classical solutions of the field equations. In the quantum theory
the space of physical configurations is also restricted by the
condition (\ref{e8}). The geometrical interpretation of this
condition has been discussed in previous work
\cite{Martin:1997cb},\cite{Martin:2001te}. We noticed that (\ref{e8}) only
restricts the values of
 $S_{s}^{r}$, which are already integral numbers from (\ref{e5}).\\

We consider now the most general map satisfying condition
(\ref{e8}). A closed one-forms $dX^{r}$ may be decomposed into the
harmonic plus exact parts: \begin{equation}
dX^{r}=M_{s}^{r}d\widehat{X}^{s}+dA^{r}
\end{equation}where $d\widehat{X}^{s}$, $s=1,2$ is a basis of
harmonic one-forms over $\Sigma$ and $dA^{r}$ are exact
one-forms. We may normalize it by choosing a
canonical basis of homology and imposing

\begin{equation} \oint_{c_{s}}d\widehat{X}^{r}=\delta_{s}^{r}. \end{equation} We have now
considered a Riemann surface with a class of equivalent canonical
basis. Condition (\ref{e5}) determines \begin{equation}\label{eq1}
M_{s}^{r}=2\pi R^{r}L_{s}^{r},\end{equation}  we rewrite $L^r_s=l_r S_s^r$ and $l_1.l_2=n$.
We now impose the condition (\ref{e8}) and obtain
\begin{equation} S_{t}^{r}\omega^{tu}S_{u}^{s}=\omega^{rs}, \end{equation} that
is, $S\in
Sp(2,Z)$. This is the most general map satisfying (\ref{e8}).
See \cite{GarciadelMoral:2008qe} for details, in particular for $n>1$.\\

 The
natural choice for $\sqrt{W(\sigma)}$ in this geometrical setting
is to consider it as the density obtained from the pull-back of the
Kh\"aler two-form on $T^{2}$. We then define \begin{equation}
\sqrt{W(\sigma)}=\frac{1}{2}\partial_{a}\widehat{X}^{r}\partial_{b}\widehat{X}^{s}\omega_{rs}.
\end{equation}

$\sqrt{W(\sigma)}$ is then invariant under the change
\begin{equation} d\widehat{X}^{r}\to S_{s}^{r}d\widehat{X}^{s}, \quad
S\in Sp(2,Z) \end{equation}

But this is just the change on the canonical basis of harmonics
one-forms when a biholomorphic map in $\Sigma$ is performed changing
the canonical basis of homology. That is, the biholomorphic (and
hence diffeomorphic) map associated to the modular transformation on
a Teichm\"uller space. We thus conclude that the theory  is
invariant not only under the diffeomorphisms generated by $\phi_{1}$
and $\phi_{2}$, homotopic to the identity, but also under the diffeomorphisms, biholomorphic
maps,
changing the canonical basis of homology by a modular transformation.\\
Having identified the modular invariance of the theory we may go
back to the general expression of $dX^{r}$, we may always consider a
canonical basis such that \begin{equation}
dX^{r}=2\pi l^rR^{r}d\widehat{X^{r}}+dA^{r}. \end{equation} the
corresponding degrees of freedom are described exactly by the
single-valued
 fields $A^{r}$. After replacing this expression in the hamiltonian (\ref{equ1}) we obtain,

\bea\label{e9}
 \begin{aligned}
H&=\int_{\Sigma}\sqrt{W}d\sigma^{1}\wedge
d\sigma^{2}[\frac{1}{2}(\frac{P_{m}}{\sqrt{W}})^{2}+\frac{1}{2}
(\frac{\Pi^{r}}{\sqrt{W}})^{2}+
\frac{1}{4}\{X^{m},X^{n}\}^{2}+\frac{1}{2}(\mathcal{D}_{r}X^{m})^{2}+\frac{1}{4}(\mathcal{F}_{rs})^{2}
\\ \nonumber &+(n^2 \textrm{Area}_{T^2}^2)+ \int_{\Sigma}\sqrt{W}\Lambda
(\mathcal{D}_{r}(\frac{\Pi_{r}}{\sqrt{W}})+\{X^{m},\frac{P_{m}}{\sqrt{W}}\})]\\& + \int_{\Sigma}
\sqrt{W} [- \overline{\Psi}\Gamma_{-} \Gamma_{r}
\mathcal{D}_{r}\Psi -\overline{\Psi}\Gamma_{-}
\Gamma_{m}\{X^{m},\Psi\}-
 \Lambda \{ \overline{\Psi}\Gamma_{-},\Psi\}]
 \end{aligned}
 \eea
where  $\mathcal{D}_r X^{m}=D_{r}X^{m} +\{A_{r},X^{m}\}$,
$\mathcal{F}_{rs}=D_{r}A_s-D_{s }A_r+ \{A_r,A_s\}$, \\
 $D_{r}=2\pi l_r
R_{r}\frac{\epsilon^{ab}}{\sqrt{W}}\partial_{a}\widehat{X}^{r}\partial_{b}$
and $P_{m}$ and $\Pi_{r}$ are the conjugate momenta to $X^{m}$ and
$A_{r}$ respectively. $\mathcal{D}_{r}$ and $\mathcal{F}_{rs}$ are
the covariant derivative and curvature of a symplectic
noncommutative theory \cite{Martin:1997cb},\cite{Boulton:2001gz}, constructed from
the symplectic structure $\frac{\epsilon^{ab}}{\sqrt{W}}$ introduced
by the central charge. The last term represents its
supersymmetric extension in terms of Majorana spinors. The physical
degrees of the theory are the $X^{m}, A_{r}, \Psi_{\alpha}$ they are
single valued fields on $\Sigma$.
\subsection{Quantum supersymmetric analysis of a single MIM2}

We are going to summarize the spectral properties of the above
hamiltonian.  The bosonic potential of the (\ref{e9}) satisfies the
following inequality \cite{Boulton:2006mm} ( in a particular gauge
condition)
\begin{equation}\label{u3}
 \begin{aligned}
&\int_{\Sigma}\sqrt{W}d\sigma^{1}\wedge d\sigma^{2}[
\frac{1}{4}\{X^{m},X^{n}\}^{2}+\frac{1}{2}(\mathcal{D}_{r}X^{m})^{2}
+\frac{1}{4}(\mathcal{F}_{rs})^{2}]\\ \nonumber & \ge
\int_{\Sigma}\sqrt{W}d\sigma^{1}\wedge
d\sigma^{2}[\frac{1}{2}(\mathcal{D}_{r}X^{m})^{2}+(\mathcal{D}_{r}A_s)^2]
 \end{aligned}
 \end{equation}
 The right hand member under regularization
 describes a harmonic oscillator potential. In particular,
 any finite dimensional truncation of the original infinite
  dimensional theory satisfies the above inequality. We consider regularizations satisfying the above inequality.
 We denote  the regularized hamiltonian of the supermembrane with
the topological restriction by $H$, its bosonic  part $H_b$ and its
fermionic potential $V_f$, then
\begin{equation} H = H_b+V_f.\end{equation}
We can define rigorously the domain of $H_b$ by means of Friederichs
extension techniques. In this domain $H_b$ is self adjoint and it
has a complete set of eigenfunctions  with eigenvalues accumulating
at infinity. The operator multiplication by $V_f$ is relatively
bounded with respect to $H_b$. Consequently  using Kato perturbation
theory  it can be shown that $H$ is self-adjoint if we choose
\begin{equation}
Dom{H}=Dom{H_b}.\end{equation}
In \cite{Boulton:2002br} it was shown that H possesses a complete set of
eigenfunctions and its spectrum is discrete, with finite
multiplicity and with only an accumulation point at infinity. An
independent proof was obtained in \cite{Boulton:2004yt} using the spectral
theorem and theorem 2 of that paper. In section 5 of \cite{Boulton:2004yt} a
rigorous proof of the Feynman formula  for the Hamiltonian of the
 supermembrane was obtained.
In distinction, the hamiltonian of the supermembrane, without the
topological restriction, although it is positive, its fermionic
potential is not bounded from below and it is not a relative
perturbation of the bosonic hamiltonian. The use of the Lie product
theorem in order to obtain the Feynman path integral is then not
justified. It is not known and completely unclear whether a Feynman
path integral formula exists for this case. In
\cite{Boulton:2006mm}  it was proved that the theory of the supermembrane with
central charges, corresponds to a nonperturbative
 quantization of a symplectic Super Yang-Mills in a confined phase and the theory
 possesses a mass gap.

 In \cite{Belhaj:2008qz}we constructed of the
supermembrane with the topological restriction on an orbifold with
$G_2$ structure that can be ultimately deformed to lead to a true G2
manifold. All the discussion of the symmetries on the Hamiltonian
was performed directly in the Feynman path integral, at the quantum
level, then valid by virtue of our previous proofs.

\section{A $U(N)$ extension of the MIM2 for arbitrary rank}
In this section we extend the algebraic symplectic structure of the
supermembrane with central charges in the L.C.G in terms of a
noncommutative product and a $U(M)$ gauge group. The main point is
to show that in such extension the original area preserving
constraint preserves the property of being first class. It is not enough to have the
symplectic structure tensor $U(M)$ in order to close the algebra of
the first class constraint. The complete expansion related to a
noncommutative associative product is needed. The noncommutative product we may introduce is
constructed with the symplectic two form already defined on the base
manifold $\Sigma$: \bea \omega_{ab}=\sqrt{W}\epsilon_{ab}, \eea
where $\sqrt{W}=\frac{n}{2}Area_{T^2}(\epsilon_{rs}\epsilon^{ab}\partial_a
\widehat{X}^r\partial_b \widehat{X}^s)$.  In this section,
in order to get a better insight on the star product, we use, without loosing generality, coordinates on the base manifold with length dimension $+1$
and define the dimensionless $\sqrt{W}$ with the area factor.
 The two-form $\omega$ define
 the area element which is preserved by the diffeomorphisms
 generated by the first class constraint of the supermembrane theory
 in the Light Cone Gauge, which are homotopic to the
 identity, and by the $SL(2,Z)$ group of large diffeomorphisms discussed in section
 2.
 The two-form is closed and nondegenerate over $\Sigma$. By Darboux
 theorem one can choose coordinates on an open set $\mathfrak{N}$ in
 $\Sigma$  in a way that $\sqrt{W}$ becomes constant on $\mathfrak{N}$.
However this property cannot be extended to the whole compact
manifold $\Sigma$. The noncommutative theory must be globally constructed
from a non-constant symplectic $\omega$. The construction of such
noncommutative theories, for symplectic manifolds
was performed in \cite{Fedosov:1996fu,Fedosov:1994zz}. The general construction for Poisson
manifolds was obtained in \cite{Kontsevich:1997vb}. \newline

The hamiltonian we propose in this section is not related to a Seiberg-Witten limit
of String Theory \cite{Seiberg:1999vs} in which one obtains a noncommutative theory with constant $B$-field.
 \subsection{The non-abelian hamiltonian}
 We now extend the above construction and consider the tensor
 product of the Weyl algebra bundle times the enveloping algebra of
 $U(M)$. It may be constructed in terms of the Weyl-algebra
 generators $T_A$ introduced in the previous section, with the
 inclusion of the identity associated to $A=(0,0)$. This complete
 set of generators determine an associative algebra under matrix
 multiplication. The inclusion of the identity allows to realize
 the generators of the $U(M)$ in terms of $T_A$
 matrices, with $A=(a_1,a_2)$ and $a_1,a_2=-(M-1),\dots,0\dots M-1$. All the
 properties of the Fedosov construction remain valid, in particular
 the associativity of the star product. It is also valid the
 Trace property.
In order to
construct the hamiltonian of the theory we consider the following
connection on the Weyl bundle \cite{Martin:2001zv} \bea
\mathcal{D}\diamond=\frac{i}{h}[G_r
e^r,\diamond]_{\circ}+\frac{i}{h}[\mathcal{A}_r e^r, \diamond]_{\circ}\eea where
$G_r, \mathcal{A}_r\in C^{\infty} (W_{Abelian})$, $\sigma G_r=
\delta_{rs}X_h^s$ and $X_h^s= 2 \pi R_s l_s \widehat{X}^s$.
It corresponds to the harmonic sector of the map to the compact sector of the target space.
$\sigma \mathcal{A}_r=A_r$ using the notation of
section 2, $e^r=\partial_a \widehat{X}^r d\sigma^a$. Its curvature
is given by \bea\label{u2} \Omega=\frac{i}{2h}[G,G]_{\circ}+\frac{i}{h}[G,\gamma]_{\circ}
+\frac{i}{2h}[\gamma,\gamma]_{\circ},\quad \gamma=\mathcal{A}_r e^r \eea

We now consider $(X^m,P_m), (A_r,\Pi^r)$ the
canonical conjugate pairs as well as the spinor fields $\Psi$
lifted to the quantum algebra $W_{abelian}\in C^{\infty}(W)$. The
constraint is then defined as \bea
\phi(\sigma,\xi,h)\equiv &=\mathcal{D}_r
\frac{\Pi^{rA}}{\sqrt{W}}T_A+ \frac{i}{h}(X^{mB}\circ
\frac{P_m^C}{\sqrt{W}}-\frac{P_m^B}{\sqrt{W}}\circ X^{C}_m)T_B T_C +\frac{i}{h}[\overline{\Psi}\Gamma_{-},\Psi]_{\circ}.
\eea with \bea
\mathcal{D}_r
\frac{\Pi^{rA}}{\sqrt{W}}T_A& = \frac{i}{h}[G_r,\frac{\Pi^{rA}}{\sqrt{W}}]_{\circ}T_A+
\frac{i}{h}(A^{B}_r\circ \frac{\Pi^{rC}}{\sqrt{W}}-\frac{\Pi^{rB}}{\sqrt{W}}\circ
A^{C}_r)T_B T_C.\eea
 We notice that the first two terms of the commutator
\bea [X^{m},\frac{P_m}{\sqrt{W}}]_{\circ}=X^{mB}\frac{P_{m}^C}{\sqrt{W}} f_{BC}^E T_E
+(-i \frac{h}{2})\{ X^{mB},\frac{P_m^C}{\sqrt{W}}\}d_{BC}^E T_E+ O((h\omega)^2) \eea are
the terms which we considered in the previous section as extensions of the algebraic
structure of the supermembrane in the Light Cone Gauge. The
additional terms arising from the noncommutative product, ensuring
an associative product, are relevant in order to close
the constraint algebra. In fact using the trace properties
$\phi\in W_{abelian}$ is a first class constraint generating a gauge
transformation which is a deformation of the original are preserving
diffeomorphisms.

The projection of $\Omega$ in (\ref{u2}) has the expression
\cite{Martin:2001zv}

\bea\begin{aligned}
\sigma\Omega=&-\omega+\mathcal{F}-\frac{h^2}{96}
(R_{bcda}\big(D_{\widehat{b}}D_{\widehat{c}}D_{\widehat{d}})A_m
-\frac{1}{4}R_{\widehat{b}\widehat{c}\widehat{d}p}\epsilon^{pq}D_q
A_m\big)\epsilon^{b\widehat{b}}\epsilon^{c\widehat{c}}\epsilon^{d\widehat{d}}e^a\wedge
e^m \\ \nonumber
&-\frac{h^2}{96.8}R_{bcda}R_{\widehat{b}\widehat{c}\widehat{d}m}
\epsilon^{b\widehat{b}}\epsilon^{c\widehat{c}}\epsilon^{d\widehat{d}}e^a\wedge
e^m+O(h^3)\dots.\end{aligned}\eea

and
\bea \mathcal{F}=\frac{1}{2}e^r\wedge e^s (D_r A_s -D_s A_r
+\frac{i}h{\{A_r,A_s\}}_{\ast}), \eea
with $\omega=\frac{1}{2h}\sqrt{w}\epsilon_{ab}d\sigma^a \wedge d\sigma^b$,
  $D_r,D_s$ are the ones defined in section 2. We remark
that $G_r$ is the lifting to the Weyl algebra of the harmonic
$\widehat{X}_r$ of section 2, and $\mathcal{A}_r$ is the lifting of
$A_r$. The $O(h^2)$ depend explicitly on the Riemann tensor of the symplectic
connection,which itself depends on the symplectic
two-form introduced by the central charge.
The $O(h)$ terms are necessary in order to close the constraint algebra.
The star product formula involve
the covariant derivatives constructed from the symplectic connection as well
as terms involving the Riemann
tensor of symplectic connection, which are absent in the Moyal product.

The hamiltonian of the theory for  $M$ multiple parallel
M2-branes with $U(N)$ gauge group is then \bea\begin{aligned}
Tr\int_{\Sigma}\mathcal{H}=Tr\int_{\Sigma}\sqrt{W}&\Bigg[\frac{1}{2}(\frac{P^m}{\sqrt{W}})^2
+\frac{1}{2}(\frac{\Pi^r}{\sqrt{W}})^2
+\frac{1}{2h^2}(\{X_h^r,
X^m\}_{\ast}+\{A^r,X^m\}_{\ast})^2\\ \nonumber &
+\frac{1}{4h^2}\{X^m,X^n\}^2_{\ast}
+\frac{1}{2}(\mathcal{F}_{rs}-\omega_{rs})(\mathcal{F}^{rs}-\omega^{rs})\\ \nonumber &
- \frac{i}{h}\overline{\Psi}\Gamma_{-}\Gamma_{r} (\{X_h^r,
\Psi\}_{\ast}+\{A^r,\Psi\}_{\ast}) -\frac{i}{h}\overline{\Psi}\Gamma_{-}
\Gamma_{m}\{X^{m},\Psi\}_{\ast}\Bigg],\end{aligned}\eea where the term
$\{{X}^r_h,X^m\}_{\ast}+\{A^r,X^m\}_{\ast}=\delta^{rs}\mathcal{D}_s X^m
+O(h)$ in the notation of section 2.
The hamiltonian is subject to the first class constraint
\bea
\phi\equiv \{X_h^r, \frac{\Pi^r}{\sqrt{W}}\}_{\ast}+\{ A_r, \frac{\Pi^r}{\sqrt{W}}\}_{\ast}
+\{X^m,\frac{\Pi^r}{\sqrt{W}}\}_{\ast}-\{\overline{\Psi}\Gamma_{-},\Psi\}_{\ast}=0
\eea
The first terms in the star product expansion are
\bea
\phi\equiv \mathcal{D}_r\frac{\Pi^r}{\sqrt{W}}+\{X^m, \frac{\Pi^r}{\sqrt{W}}\}-\{\overline{\Psi}\Gamma_{-},\Psi\}_{\ast}+ O(h)
\eea
where $<,>_{\ast}$ has been normalized in a way to be a deformation of $\{,\}$ the symplectic bracket of the supermembrane in the L.C.G.
In the notation of section 2, the fields are now $u(M)$ valued.
 An explicit expression for $O(h)$,
the first terms with the explicit dependence were found in \cite{Martin:2001zv}, for example,
if we make manifest the dimensional dependence of the star-product we can realize
 that the parameter $[h]=n. Area_{T^2}$, $n$ is the wrapping number. In fact due to the minimal immersion map there
  is a local bijection between the coordinates in the base manifold and those in the
  compact part of the target space:
\bea
\int_{\Sigma}d^2\sigma\sqrt{W} = \int_{\Sigma} dX^1_h dX^2_h
\eea
with $X_h^r= 2\pi R^rl^r\widehat{X}^r$.
The star-product  is explicitly given by
\bea
\begin{aligned}
\frac{i}{h}\{f,g \}^a_{\ast}= & \frac{i}{h}f^b g^c f_{bc}^a + \{f^b,g^c\}d_{bc}^a+ O(h)
\\ \nonumber &= \frac{i}{n Area_{T^2}}f^b g^c f_{bc}^a+\{f^b,g^c\}d_{bc}^a+ O(n Area_{T^2})
\end{aligned}
\eea

where $\{f^b,g^c\}=\epsilon^{rs}D_r f^b D_s g^c$. $D_r$ was defined in section 2.
The factor $\frac{1}{h}$ ensures that this formalism is a
 nonabelian extension of the abelian MIM2-brane,  since for the abelian case
$f_{bc}^a$ vanishes, $d_{bc}^a=1$, and the algebra closes exactly with the ordinary symplectic
bracket corresponding to a single M2 action without further contributions.

The associated action to this hamiltonian \cite{GarciadelMoral:2009uf} is invariant under the following
supersymmetric transformations with parameter $\epsilon=\Gamma_-\Gamma_+ \epsilon$
\bea
\begin{aligned}
&\delta A_M=\delta A_M^B T_B=\overline{\epsilon }\Gamma_M \Psi^B T_B \quad M=r,m\\ \nonumber &
\delta A_0 =\delta A_0^B T_B= -\overline{\epsilon}\Psi^B T_B \\ \nonumber &
\delta \Psi= \delta \Psi^B T_B= \frac{1}{4}\Gamma_{+}\Omega_{MN}^B \Gamma^{MN}\epsilon T_B
+ \frac{1}{2}\Gamma_{+}\Omega_{0M}^{B}\Gamma^M \epsilon T_B
\end{aligned}
\eea
These transformations are a $U(N)$ extension of the SUSY transformations for the supermembrane in the LCG found by
\cite{deWit:1988ig, Bergshoeff:1987qx} and they preserve $\mathcal{N}=8$ supersymmetry.
The invariance of the action arises in a similar way as it does for Super Yang-Mills.

\subsection{Decoupling limit}
The mass square operator may be written as: \bea -mass^2=\int
(\frac{1}{2}d\widehat{x}^r\wedge
d\widehat{x}^s\epsilon_{rs})[\frac{1}{2}(\frac{P}{\sqrt{W}})^2
+\frac{1}{2}(\frac{\Pi}{\sqrt{W}})^2+(T Area_{T^2}^2)(V_B+V_F))]
\eea where $V_B$ and $ V_F$ are the bosonic and fermionic potentials
of the Hamiltonian. The scale of the theory is then
$T.n.\textrm{Area}_T^2$. The measure of integration reduces to the
dimensionless \newline
$\frac{1}{2}d\widehat{x}^r\wedge
d\widehat{x}^s\epsilon_{rs}$. The conjugate momenta have mass
dimension $+1$, and the corresponding configuration variable mass
dimension $-1$. $T$ has mass dimension $+3$. On the other hand,
 by considering the contribution to Yang-Mills arising from the first term in the
above expansion of the star product and by taking canonical
dimensions for the conjugate pairs we get for the coupling constant

\bea
g_{YM}=\frac{1}{T_{M2}^{1/2}. n. Area_{T^2}}.
\eea
It has dimension of $mass^{1/2}$. It represents the
coupling constant of the first term in the star-product expansion.
We assume that the compactification radii  is $R_i>>l_p$
but with the theory still defined at high energies.
For a fixed tension and winding number $n$,
the only relevant contribution in the star product  at low energies is the $U(M)$ commutator
since the natural length is much larger larger that the effective radii $R_{eff}=n^{1/2}\sqrt{R_1R_2}$.
This is the decoupling limit of the theory since the Yang Mills field strength becomes the coupling constant of the theory. The $g_{YM}$ is very large in this phase and the theory is in the IR phase.
It corresponds to have a description of  M multiple MIM2-branes as point-like particles, representing $M$ the number of supermembranes.
As we raise the energy the $g_{YM}$ coupling constant gets
weaker and for energies high enough, comparable with the natural scale of a MIM2-brane with an effective area of ($n.Area_{T^2}$),
the oscillation and vibrational modes containing the gauge but also gravity interactions between the supermembranes
are no longer negligible so the full star- expansion has to be considered.
All terms associated to the supermembrane symplectic structure of the star-bracket contribute while the ordinary SYM contribution vanishes.
The point-like particle picture is no longer valid, and it is substituted for that of an extended $(2+1)$D object
and the gauge and gravity contributions are strongly coupled.
One can define formally and effective physical coupling constant for the ordinary $F_{\mu\nu}$ field strength which it would correspond to
$\Lambda=M. g_{YM}$ with $M$ representing the number of supermembranes and then one can try to obtain the 't Hooft coupling expansion in the large M.
In this picture however one should take care on the limit.
By keeping $\Lambda$  fixed with $M$ going to infinity, for a fixed tension and a fixed compactification radii,
 one has to consider the wrapping number $n$ also going to infinity.
But $n.Area$ is the order parameter that would also go multiplied by $M$ in the expansion so one enters "faster" in the strong correlated
limit where the rest of the terms of the star-product expansion cannot be neglected, moreover, from a physical point of view
$n. area_{T^2}$ is related to the size of the MIM2 as an extended object and it cannot be larger than the present energy bounds we have, otherwise
it would be in contradiction with our point-particle description at low scales.
In order to perform a more accurate analysis one should be working
with the nonabelian extension of the MIM2 for 4D noncompact,-it will be considered elsewhere- however we believe that the qualitative arguments
presented here should remain valid also in that case.

\section{Discussion and Conclusions}
We have obtained a $N=8$ nonabelian  U(M) formulation of the
 minimally immersed supermembrane for arbitrary number of colors $M$ with
 all the symmetries of the supermembrane, in the LCG.
This corresponds to the M-theory
dual of the Nonabelian formulation of a \emph{bundle} of Dirac-Born-Infeld
representing a D2-branes-D0 bound state. It is the first time that
a nonabelian gauge theory can be directly obtained from a full-fledged  sector of M-theory
element, so far restricted to String theory: Heterotics and
 Dp-branes in type II theories. This opens a new interesting
window for models in phenomenology.
At energies of the order of the compactification scale, the theory has the gauge and gravity sector strongly coupled.
It describes all of the oscillations modes of the multiple parallel M2-branes minimally immersed.
At low energies the theory enters in a decoupling regime and the physics is then described
 by a $\mathcal{N}=8$ SYM theory of point-like particles in the IR phase. We then expect to describe correctly many aspects of
phenomenology when realistic gauge groups will be considered.  From the point of view of the target space the
theory has N=1 susy in 9D flat-dimensions. In \cite{GarciadelMoral:2007xj}
a N=1 target space, D=4 formulation of a single
supermembrane minimally immersed together with a number of
interesting phenomenological properties were found. Moreover in \cite{Belhaj:2008qz} a
formulation of the supermembrane minimally immersed on a G2 manifold
was also obtained. Its quantum supersymmetric spectrum is also
purely discrete. The analysis in 4D can be also extended to the
nonabelian case following the lines shown in this paper,
allowing to obtain models with reduced number of target and worldvolume supersymmetries.

\section{Acknowledgements}
The work of MPGM has been partially supported by the grants MICINN-09/FPA2009-07122,
FICYT-09/IB09-069 and the MEC-DGI Consolider CSD2007-00042. The work of AR is partially  by
PROSUL, under contract CNPq 490134/2006-08.

\bibliographystyle{JHEP}

\bibliography{lista1}


\end{document}